\documentclass[
 aip,
 amsmath,amssymb,
 reprint,%
]{revtex4-1}

\usepackage{caption}
\usepackage{subcaption}
\newcommand{\circled}[1]{\raisebox{.5pt}{\textcircled{\raisebox{-.9pt} {#1}}}}

\usepackage{graphicx}% Include figure files
\usepackage{dcolumn}% Align table columns on decimal point

\usepackage{bm}% bold math
\usepackage[utf8]{inputenc}
\usepackage[T1]{fontenc}
\usepackage{mathptmx}
\usepackage{etoolbox}

\makeatletter
\def\@email#1#2{%
 \endgroup
 \patchcmd{\titleblock@produce}
 {\frontmatter@RRAPformat}
 {\frontmatter@RRAPformat{\produce@RRAP{*#1\href{mailto:#2}{#2}}}\frontmatter@RRAPformat}
 {}{}
}%
\makeatother
\begin{document}

\preprint{AIP/123-QED}

\title[broadband shot-to-shot transient absorption anisotropy]{Broadband shot-to-shot transient absorption anisotropy}
% Force line breaks with \\
\author{Maximilian Binzer}
 \affiliation{Technical University of Munich, TUM School of Natural Sciences, Department of Chemistry, Professorship of Dynamic Spectroscopy, 85748 Garching, Germany}%Lines break automatically or can be forced with \\

\author{Franti\v{s}ek \v{S}anda}%
\affiliation{ 
Institute of Physics, Faculty of Mathematics and Physics, Charles University, Prague 12116, Czechia
}%

\author{Lars Mewes}
 \affiliation{Technical University of Munich, TUM School of Natural Sciences, Department of Chemistry, Professorship of Dynamic Spectroscopy, 85748 Garching, Germany}
 \affiliation{WSL-Institut für Schnee- und Lawinenforschung SLF}
 
\author{Erling Thyrhaug}
 \affiliation{Technical University of Munich, TUM School of Natural Sciences, Department of Chemistry, Professorship of Dynamic Spectroscopy, 85748 Garching, Germany}

\author{J\"{u}rgen Hauer}
\email{juergen.hauer@tum.de}
 \affiliation{Technical University of Munich, TUM School of Natural Sciences, Department of Chemistry, Professorship of Dynamic Spectroscopy, 85748 Garching, Germany}

\date{\today}% It is always \today, today,
 % but any date may be explicitly specified

\begin{abstract}
Transient absorption (TA) is the most widespread method to follow ultrafast dynamics in molecules and materials. The related method of TA anisotropy (TAA) reports on the ultrafast reorientation dynamics of transition dipole moments, reporting on phenomena ranging from electronic dephasing to orientational diffusion. While these are fundamental aspects complementary to TA, TAA is generally less widely used. The main reason is that TAA signals are usually not measured directly but are retrieved from two consecutive TA measurements with parallel (${\mathcal R}_\parallel$) and perpendicular (${\mathcal R}_\perp$) polarization of pump and probe pulses. This means that even minor systematic errors in these measurements lead to drastic changes in the TAA signal. In this work, the authors demonstrate alternating shot-to-shot detection of ${\mathcal R}_\parallel$ and ${\mathcal R}_\perp$, minimizing systematic errors due to laser fluctuations. The employed broadband detection lets us discuss effects dependent on detection wavelength in the ultrafast anisotropy decay of 2,3-Naphthalocyanin, a system previously scrutinized by David Jonas and co-workers. In particular we compare timescales of population relaxation and decoherence and supported the proposals for isotropic type of relaxation in square symmetric molecules. 
\end{abstract}

\maketitle

\section{\label{sec:int} Introduction}

Ultrafast spectroscopies with femtosecond optical pulses became widely used in biology, chemistry, and physics research.\cite{Maiuri2019} 

Among these non-linear techniques, ultrafast transient absorption (TA) based on femtosecond excitation and the broad detection windows achievable with supercontinuum whitelight detection stands out. While broadband detected TA is straight-forward to implement, deducing the ultrafast kinetics from these experiments can pose a problem, as there may be substantial overlap of transient signals contributions that leaves the spectra hard to disentangle.
Besides advances in data analysis, \cite{Stokkum2004} experiments capable of disentangling transient signals in congested spectral regions are needed. 
One such approach to this problem was the introduction of 2D electronic spectroscopy, where both the excitation and detection frequency dependence of the signal are fully resolved. \cite{Jonas2003, Lepetit1996, Cho2008} While this approach leads to significant de-congestion of the optical spectra, 2D spectroscopies represent a significant increase in experimental complexity relative to TA. 
At a level of experimental complexity comparable to TA, transient absorption anisotropy (TAA) is one approach that increases resolution of ultrafast signals by their polarization dependence.\cite{Albrecht1961,Hybl2002,Farrow2008,Xu2023,Xu2024} This gives information about the transition dipole moments, their time-dependent relative orientation and fluctuations, and - via dephasing dynamics - insights on solute-solvent interaction. \cite{Mewes2020,Norden1977,Thyrhaug2013,Wallin2005} 
TAA provides this important information in an experiment that is relatively simple compared to the more evolved polarization-controlled electronic 2D-spectroscopy.\cite{Tiwari2012,Thyrhaug2018,Mancal2012}
Furthermore, TAA can be used to determine energy transfer between spectrally identical molecules, which is especially relevant for biological light harvesting complexes. \cite{Martinsson2000,Jonas1996} While the origin of dephasing behavior is often difficult to determine in TA, transient absorption anisotropy allows direct measurement of electronic dephasing. \cite{Galli1993}

The anisotropy $r$ of a system cannot be measured directly, but needs to be calculated from two independent measurements: one with the field polarization of excitation and probe pulses parallell (${\mathcal R}_\parallel$) and one with the fields polarizations orthogonal (${\mathcal R}_\perp$). From these measurements one can calculate the TAA from the standard expression 
$r = \frac{{\mathcal R}_\parallel - {\mathcal R}_\perp}{{\mathcal R}_\parallel+2 {\mathcal R}_\perp}$. 
In practice, these measurements are typically done consecutively, with the pump pulse polarization being rotated 90 degrees by use of a half $\lambda$ plate between the measurements. As for any signal derived from two separate measurements, long-term drifts in the laser intensity are problematic.

With the advent of high repetition rate laser systems and cameras capable of detecting broadband spectra in the regime of up to 100 kHz, data acquisition times - as well as signal-to-noise ratios - improved dramatically. \cite{Donaldson2023}
In this work, we utilize these technological advances in laser source and detector technology and combine them with a Sagnac-Interferometer.\cite{Courtney2014} The general purpose of an interferometer is to generate two replicas of the entering pulse with precisely controllable delay within the produced pulse pair. In the experiment described here, the delay within the pulse pair (as well as the phase stability between the replicas) are of no importance. We rather use the fact that the resulting pulse pair is collinear and - most importantly - the polarization of each replica can be modified individually. In combination with optical choppers and fast detection, this allows us to record ${\mathcal R}_\perp$ and ${\mathcal R}_\parallel$ on a shot-to-shot basis, eliminating drift-related problems in conventional (sequential) approaches. The ability to accommodate separate polarizing optics in a single-pass fashion as well as chopper wheels in both interferometer arms is the main advantage of the Sagnac over more widespread interferometer designs such as Michelson. \cite{Giacomo1987,Zetie2000,Arianfard2023}
 The Sagnac-Interferometer employed in this work allows for the creation of a sequence of two pulses with tunable femto\-second delay and arbitrary relative polarization. It can be implemented as a simple add-on to an existing TA setup, placed in the pump arm of the experiment. This amounts to a straightforward implementation of a transient absorption anisotropy (TAA) experiment as demonstrated in this work. 

\section{\label{sec:exp} Experimental Design and Results}

\begin{figure*}
\caption{Experimental Setup}
	\includegraphics{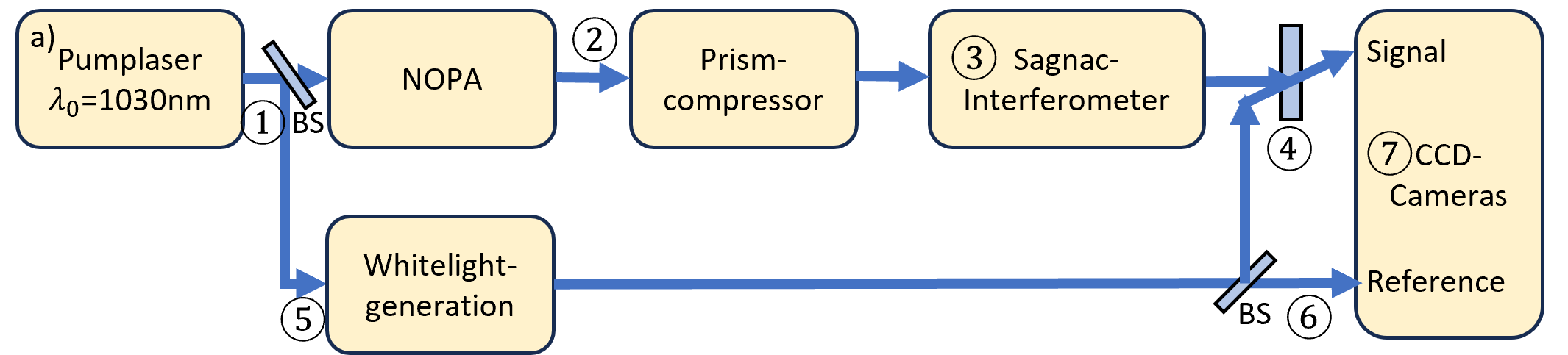}	
	\label{fig:setup}

	\begin{minipage}{0.55\textwidth}%not 0,.5 or sagnac too close to other graph
		\includegraphics{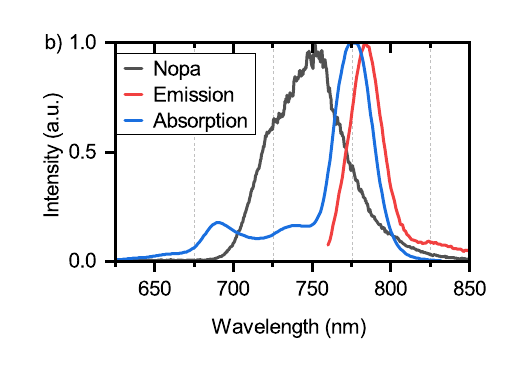}
	\end{minipage}
	\begin{minipage}{0.4\textwidth}
		\includegraphics{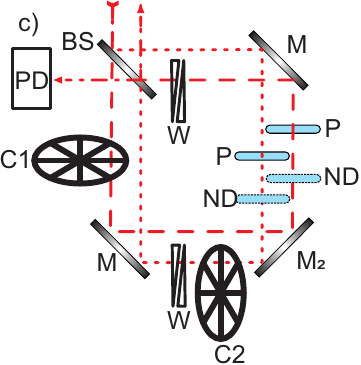}

	\end{minipage}

	\includegraphics{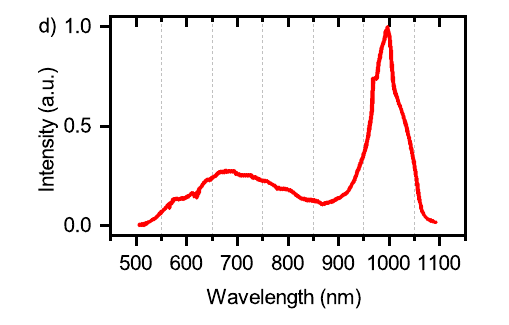}
	\includegraphics[trim=0 -1.8cm 0 0]{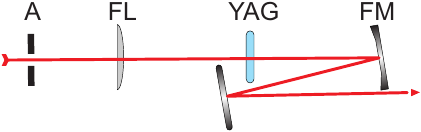}

 \caption{(a) Sketch of Setup, (b)\circled{2} NOPA pump spectrum and overlapped absorption and emission of the 2,3-Naphthalocyanin, (c) \circled{3} Sagnac-Interferometer,(d) \circled{5} Whitelightgeneration}
\end{figure*}

We show a sketch of the experimental setup in Figure \ref{fig:setup}. The 1030\,nm (230 fs), 75\,µJ per pulse output of a Yb:KGW amplified laser, equipped with a pulse picker for repetition rate control, (Pharos PH2, LightConversion UAB.) is split into pump- and probe-paths on a 80/20 beamsplitter shown at \circled{1} in Figure \ref{fig:setup}. 
In the pump arm, the 60 µJ/1030 nm laser pulses are frequency converted and spectrally broadened using a non-collinear optical parametric amplifier (NOPA Rainbow \cite{Schriever2008}) at \circled{2} resulting in 3µJ of power remaining. A representative visible-range pulse spectrum is shown in Figure \ref{fig:setup}(b). The broadband pulses are subsequently pre-compressed to compensate for spectral chirp induced by the optical components using a fused silica prism-pair compressor. In this configuration, the pump-pulse central wavelength is freely tuneable in the 400-900 nm range, and are readily compressed to sub-20 fs. For the experiment in this work, the pulse duration was 18 fs at the sample position as determined by an autocorrelation measurement. \cite{Kozma2004}
The resulting beam is routed to a Sagnac-Interferometer at \circled{3}, where the incoming pulse is split and then recombined to yield a collinear pulse pair with controllable relative linear polarization. Details are given in the next subsection. The output of the Sagnac-Interferometer serves as a pump pulse in a TA experiment with the sample at position \circled{4}. 

In the probe arm the 15 µJ/1030 nm pulses are focused into a 3\,mm YAG crystal at \circled{5} in order to generate the supercontinuum whitelight used for signal detection. The resulting broadband spectrum covers the range between 500 and 1000\,nm and is shown in figure \ref{fig:setup} (d)). An aperature (A) is used to control the spotsize on the focusing lens (FL, with focal length $f$=75 mm). The resulting whitelight is collimated by a concave mirror with $f$=-100\,mm. Before reaching the sample, a small fraction of the supercontinuum probe light is reflected on a quartz plate at \circled{6}. The light transmitted through the quartz plate is overlapped with the pump in the sample before frequency-dispersed detection in a prism spectrograph, equipped with a 50 KHz read-out rate CCD-camera (Entwicklungsbüro Stresing)\circled{7}. The smaller fraction of the probe beam does not pass the sample but is detected on an additional camera to provide a reference for a balanced detection scheme. \cite{Brazard2015}

We employ these experimental developments to revisit excitation relaxation in a square-symmetric molecule, the 2,3-Naphthalocyanin. Observing TA and TAA at several detection frequencies we resolve the relaxation dynamics and discuss the relation between decoherence and population relaxation rates in square symmetric molecules.

\subsection{\label{sec:sagnac}Sagnac-Interferometer}
We generate a pair of collinear femtosecond broadband pulses in the pump arm of the instrument using a Sagnac-Interferometer (see figure \ref{fig:setup} \circled{4}). This interferometer design is advantageous since both pump pulses pass through the same optical elements and amount of material, thus minimizing potential differences in pulse properties within a pair. 
As a modification of the standard Sagnac implementation we displace the routing mirror opposite the beamsplitter, \textbf{$M_2$} in Figure \ref{fig:setup}(c), which distorts the interferometer geometry with respect to the conventional perfect square. This allows us to spatially displace the clockwise- (short dashed) and counter-clockwise (long dashed) interferometer paths relative to each other by a distance of up to the diameter of the mirrors. This is shown in more detail in figure \ref{fig:setup}(c). The incoming beam arrives on the top left and is split into two seperate beam paths. The dashed beam path is first transmitted through the beamsplitter(\textbf{BS},Layertec 100937), reflected three times at mirrors(\textbf{M}), and then reflected at the beam splitter. The short dashed beam path is first reflected and then transmitted through the beamsplitter at the end of its propagation. 

With the two interferometer arms spatially separated, they can be manipulated individually by \textit{e.g.} polarizing optics and choppers.
The relative timing difference between the interferometer arms is controlled by pairs of fused silica wedges (\textbf{W},2° Fused Silica), which allow fine control of the optical path lengths in the two arms. For the anisotropy measurements the timing difference is zero, which can be set interferometrically or by measuring the TA signal with pump pulses from each arm and matching the delay time at which the coherent artefact occurs. \cite{Dobryakov2005}
The pulse power in each arm is adjusted by using variable attenuators (Thorlabs), while the pulse polarization set by half-$\lambda$ plates (\textbf{P}) (B. Halle Nachfl. GmbH, Art.Nr. BHN 2016.0070.0008).

\begin{figure}[ht] 
 \centering
 \includegraphics{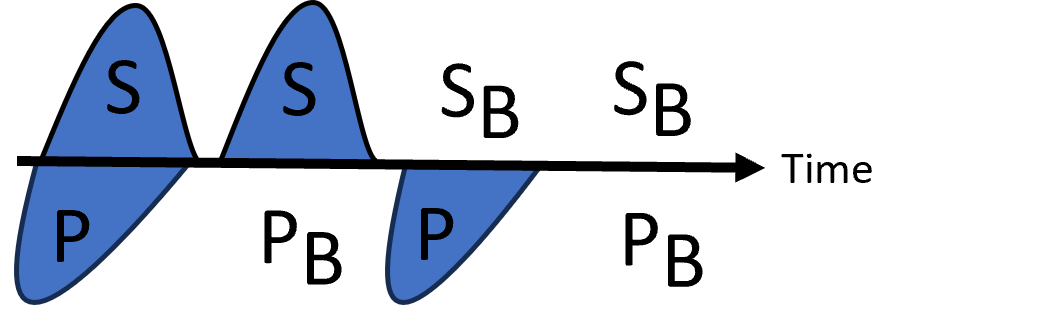}
 \caption*{(a)Pulsetrain with the four polarization combinations}

 \includegraphics{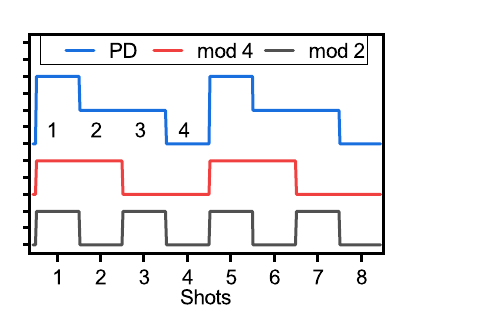}
 \caption{(b) The Sagnac-Interferometer is used to generate a four-pulse sequence with different polarizations. The chopper in the arm of the interferometer with $s$-polarized light transmits two pulses in a 5 kHz input sequence and blocks the next two. The other chopper in the arm with $p$-polarized light blocks every second shot. This leads to a sequence starting with two simultaneous and orthogonally polarized pulses, followed by single pulses with $s$ and $p$ polarization respectivly. In the last event of the sequence, both interferometer arms are blocked. The lower panel shows the chopping sequence and the schematic readout of the photodiode in Figure \ref{fig:setup} for both paths open and each arm seperately.}
 \label{fig:pulsetrain} 
\end{figure}

The choppers (\textbf{C}, New Focus Optical Chopper 3501) are used to obtain a different chopping frequency in each arm. The aim is to deliver a pulse sequence with which the TAA signal can be measured on a shot-to-shot basis. The arm in which the polarization has been set to $p$ (for later interaction with the sample), the chopper operates at half the laser repetition rate, leading to a 2.5 kHz repetition rate in this experiment (see Figure \ref{fig:pulsetrain}). In the arm with $s$-polarization (orthogonal to the laser table), the chopper operates at 1.25 kHz - one fourth of the repetition rate. This leads to a repeated sequence of 1) two pulses ($s$ and $p$), 2) one pulse ($s$), 3) one pulse ($p$) and 4) no pulse reaching the sample position. TAA is calculated from the signals obtained with single excitation pulses (2 and 3) and the no-pulse scenario (4). The two-pulse event (1) is so far disregarded. 

Intensity fluctuations and possible artefacts due to chopping mistakes are monitored by a photodiode (Thorlabs DET10A2), see Figure \ref{fig:setup}(c). The photodiode readout value was saved on a freely programmable pixel of the CCD detector (Entwicklungsbüro Stresing). This allows to correlate each laser spectrum with the corresponding chopper states (see lower panel in Figure \ref{fig:pulsetrain}).
Similar chopping schemes were employed previously for noise suppression\cite{Augulis2011} and shot-to-shot detection in 2D-ES.\cite{Son2017} 
The repetition rate of used in this experiment is limited by the optical choppers, however this limiation could be circumvented by the use of \textit{e.g.} acousto optic modulators (AOMs) \cite{Tekavec2007,Mueller2019,Wituschek2020} if a higher repetition rate was desired. More importantly, AOMs lead to additional chirp on the excitation pulses, which would worsen the temporal resolution of the experiment. \cite{Tekavec2007}

The intensity of the subpulses from the different arms must be equal to obtain a meaningful TAA signal. While for TA measurements a misidentification of the chopper state can be corrected in data analysis, correcting a TAA measurements in post processing is prone to error. We monitor the intensity of the entire four-pulse sequence in Figure \ref{fig:pulsetrain} by a photodiode, PD in Figure \ref{fig:setup}(c). 
We retrieve the intensity of the two subpulses used for pumping the sample by associating the PD readout (blue in Figure \ref{fig:pulsetrain}(b)), to the chopper phase in both arms (red and black).

The pump-probe delay is set by a mechanical stage (Newport XMS50). The sample is stored in a cuvette on a stage that is quickly oscillating on an axis orthogonal to the beam path to ensure sample exchange between laser shots. 

\begin{figure} 
	\includegraphics{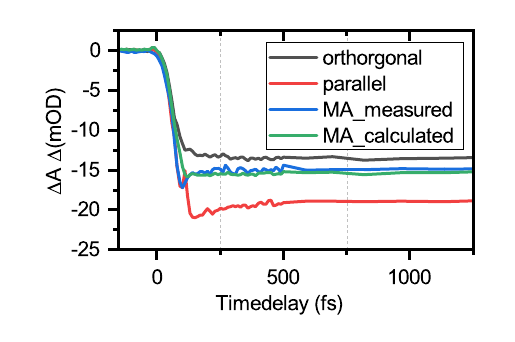}
	\caption{The parallel and orthogonal components of the transient absorption (TA) signal at 790 nm detection wavelength. The good overlap between the directly measured TA-signal in magic angle (MA) and the reconstructed MA signal, $\mathcal{R}_{MA}=(\mathcal{R}_\parallel+2\mathcal{R}_\perp)/3$, demonstrates the validity of the anisotropy measurement.}
 \label{fig:par_ort}
\end{figure}

To evaluate the correct alignment of the interferometer, the magic angle (MA) TA signal - measured with one interferometer arm blocked and the polarization between pump and probe set to 54.7° - was compared to the respective signal as retrieved from $\mathcal{R}_{\parallel}$ and $\mathcal{R}_{\perp}$ via $\mathcal{R}_{MA}=(\mathcal{R}_\parallel+2\mathcal{R}_\perp)/3$. With $\mathcal{R}$ we denote measured or calculated TA-signals with the subscript giving the relative polarization of pump and probe pulses. As seen in Fig. \ref{fig:par_ort}, the agreement between the directly measured $\mathcal{R}_{MA}$ and the TA-signal retrieved from the polarized measurements is good. 

After discussing the experimental design and characterization of the TAA spectrometer, we now turn to the calculation of the TAA-signal. As the goal is to measure shot-to-shot anisotropy, the most straightforward approach is to calculate the delay time dependent anisotropy $r(t)$ for each four-pulse sequence and then average afterwards:
\[
 r(t)= \left\langle \frac{log(\frac{S_\parallel}{S_b}) - log(\frac{S_\perp}{S_b})} {log(\frac{S_\parallel}{S_b}) +2 log(\frac{S_\perp}{S_b})} \right\rangle 
\]
$S_b$ stands for the background signal with all excitation pulses blocked, see Figure \ref{fig:pulsetrain}. $S_{\parallel,\perp}$ denote measured probe transmission signals for parallel and orthogonal pump polarization. The anisotropy signals as calculated by the equation above leads to a poor signal-to-noise ratio (data not shown). We explain this by the equation's logarithm, as small differences in signal strength ($S_\parallel$, $S_\perp$ and $S_b$) will result in large differences in $r$. Alternatively, we first average the signals to then calculate the $r$: 

\[
r(t)= \frac{log(<\frac{S_\parallel}{S_b}>) - log(<\frac{S_\perp}{S_b}>)}
{log(<\frac{S_\parallel}{S_b}>) +2 log(<\frac{S_\perp}{S_b}>)}
\]

An advantage of this approach is that we can apply singular value decomposition on the individual TA signals and filter out low-intensity components. \cite{Stokkum2004} This step enhances the signal-to-noise ratio of the TA-signals prior to calculating $r(t)$.

The sample used is 2,3-Naphthalocyanin dissolved in Dimethylsulfoxid (DMSO).
Naphthalocyanines are phthalocyanine derivates with an extra phenyl group in each arm of the chromophore. The square symmetric properties ($D_{4h}$ point group) are retained. We chose this class of molecules due to their fast anisotropy decay, as reported previously by David Jonas and co-workers. \cite{Farrow2008} 

\begin{figure}
 \includegraphics{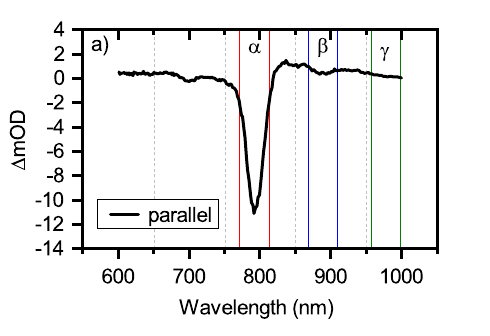}
 \includegraphics{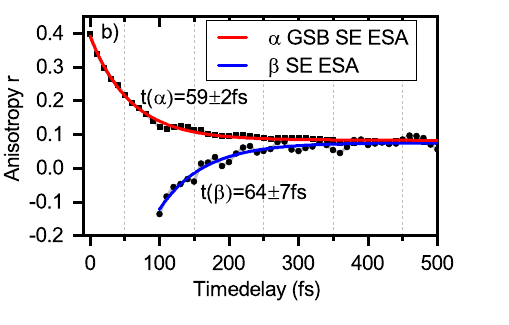}
 \includegraphics{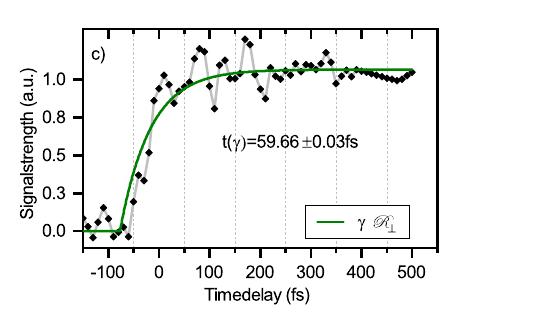}

	\caption{(a) The TA signal with parallel polarization for pump and probe. Blue and red vertical lines indicate the two spectral regions centered around 890 and 790 nm where the signal was analyzed by TAA. (b) GSB, ESE, and ESA determine the TAA signal described by the blue decay curve. Only contributions from ESA and ESE are present for the trace around 890 nm, fitted by the red curve. Both signals show the same 60 fs rise/decay time (c) In the region around 1000 nm, labeled as $\gamma$ in (a), only ESA pathways contribute. The $\mathcal{R_\perp}$ signal shows a 60 fs decay time. c) The fit to the orthogonal TA signal at $\gamma$ centered around 980 nm.}
	\label{fig:decay_mult}
\end{figure}

In figure \ref{fig:decay_mult} we show the broadband detected TA-signal at a pump-probe delay of 300 fs. 
The spectral region in the detection window around 790 nm, $\alpha$ in Figure \ref{fig:decay_mult}(a), is defined by ground state bleach (GSB), stimulated emission (SE) and excited state absorption (ESA). The corresponding anisotropy signal, shown in Figure \ref{fig:decay_mult}(b), shows an initial value of 0.4 and decays with a time constant of 60 fs. The corresponding fit is shown in red in Figure \ref{fig:decay_mult}(b). The same time constant (blue line) is found for the rise time of the dynamics around 890 nm, see region labelled with $\beta$ in Figure \ref{fig:decay_mult}(a). In this region, we expect contributions from SE and ESA pathways. In the region labelled as $\gamma$ in Figure \ref{fig:decay_mult}
(a), we expect contribution only from ESA pathways, as $\gamma$ shows overlap with neither the absorption nor the emission spectrum, see Figure \ref{fig:setup}(b). As the TA-signal strength in this region is weak, $r(t)$ is too noisy for analysis (data not shown). Instead, we show the $\mathcal{R}_\perp$-signal in Figure \ref{fig:decay_mult}(c) and find the same 60 fs decay time. 

\section{Discussion}\label{sec:result}

In the following section, we develop a model for the ultrafast anisotropy decay $r(t)$ of 2,3-Naphthalocyanin based on a standard response function formalism. \cite{Mukamel2009} 
Within this framework, we will explain the mono-exponential (60 fs) anisotropy decay dynamics and the different initial anisotropy values in spectral regions discussed in Figure \ref{fig:decay_mult}. We will also describe why the 60 fs component is absent in the $\mathcal{R}_{MA}$ signal (see Figure \ref{fig:par_ort}) but present for $\mathcal{R}_\perp$ (see Figure \ref{fig:decay_mult}(c) and $\mathcal{R}_\parallel $.

The transient absorption spectra are defined by transitions between three excitonic manifolds of a molecule. A model of two electrons in a square 2D box has been proposed to allow for a basic understanding of the ultrafast dynamics of square symmetric molecules. It was argued that the anisotropy dynamics of porphyrins and an electron in a 2D-box are the same,\cite{Qian2003} (despite known differences in the excited state symmetries of a 2D-box and naphtalocyanines\cite {Orti1993}). After an ad hoc adjustment was made for the second excitonic manifold, this approach was successful in modeling the anisotropy decay of 2,3-Naphthalocyanin bis(trihexylsilyoxide). \cite{Farrow2008} We adhere to the two electrons in the 2D box picture, and work in the basis introduced by Qian \textit{et al.}\cite{Qian2003} We also note that in these manifolds of degenerated states the straightforward transformation between the different (2D box \textit{vs}. porphyrin states) basis sets is possible, which only interchanges the role of the population relaxation with decoherence. 
We will demonstrate that the 2D-box model is applicable to simulate the TA and TAA signal of 2,3-Naphthalocyanin while keeping full account of both decoherence and population relaxation dynamics, thus avoiding the ad-hoc corrections about the orbital symmetries made by Farrow et al. \cite{Farrow2008}

The TA-spectra are defined by transitions between three excitonic manifolds of a square symmetric molecule (see Supporting Information). Light-matter interactions induce transitions between the ground- and the first excited manifold as well as between the first and the second excitonic manifold..

In general, a TA experiment interrogates the relaxation dynamics within the manifold of two degenerated singly excited states $|1E_{ux}\rangle$ and $|1E_{uy}\rangle$. Within this manifold, two relaxation pathways are considered. The excitonic coherence between the first excited states dephases due to random fluctuations, following a decay function of the type ${{\mathcal I}(t) \equiv\mathcal I}_{xy}(t)={\mathcal I}^*_{yx}(t)=e^{-\Gamma_{{\rm dech}} t}$ in the Markov approximation. $\Gamma_{{\rm dech}}$ stands for the decoherence rate. At longer timescales, population relaxation between the two excitons is expected as a second relaxation pathway, described by the Greens function $G_{ij}(t)$ with $G_{xj}(t)+G_{yj}(t)=1$. When we assume symmetry between the two excitons, we obtain $G(t)\equiv G_{xx}(t) = 1-G_{yx}(t)=G_{yy}(t)=1-G_{xy}(t)$. In the Markov approximation $G(t)=(1+e^{-\Gamma_{pop} t})/2$, where $\Gamma_{pop}$ is the population relaxation rate, in other words $\Gamma_{pop} /2$ is the rate of population transfer from one of the first excited states to the other. 

Raghavan \textit{et al.} \cite{Raghavan2000} suggested that in a strict square symmetry, the population relaxation and decoherence times between degenerated states of the first excitonic manifold show the same numerical value. 
Although we do not think that this equality is a mathematically inevitable consequence of the square symmetry, it is a reasonable proposal: both processes are induced by the same low frequency vibrations. TA at different detection frequencies can help us to prove or disprove the proposed equality between $\Gamma_{pop}$ and $\Gamma_{{\rm dech}}$ experimentally.

TA is a third order perturbative response described by a rank 4 tensor, the response function. In the supporting information, we evaluate this tensor in the molecular frame $R^{ijkl}$, indices $i,j,k,l$ stand for excitation along the $x$ or $y$ -oriented transition dipole moment for a square symmetric molecule. The response function is dissected into the contributions from GSB, SE and ESA pathways as depicted by double-sided Feynman diagrams shown in SI (Fig. S1). The ESA is further summed from the four degenerated doubly excited states with $|2A_{1g} \rangle$, $ |2B_{1g} \rangle$ and two with $|2B_{2g} \rangle$ symmetry. 
Next the two lab frame polarization-controlled measurements, ${\mathcal R}_{\parallel}$ and ${\mathcal R}_{\perp}$, which are relevant for TA and TAA measurements are evaluated.

These expressions are combined in MA-TA:
\begin{equation}
 {\mathcal R}_{MA}= {\mathcal R}_{\parallel}+2{\mathcal R}_{\perp}=\sum_{ij}\frac{1}{3} R^{iijj}
\label{eq:PP}
\end{equation}
The TAA signal, also referred to as the anisotropy ratio, is given by
\begin{equation}
 r= \frac{{\mathcal R}_{\parallel}-{\mathcal R}_{\perp}}{{\mathcal R}_{\parallel}+2{\mathcal R}_{\perp}}=\frac{\sum_{ij} (-2R^{iijj}+3 R^{ijij}+3R^{ijji})}{10 \sum_{ij} R^{iijj}} 
 \label{eq:anisotropy}
\end{equation}

The ${\mathcal R}_{MA}$ signal as defined by eq. \ref{eq:PP} is proportional to an isotropic invariant $R^{iijj}$ known to be insensitive to orientational dynamics such as (orientational) diffusion \cite{Hamm2011}. It is clear that the signal is insensitive to excitonic dephasing dynamics between $x$- and $y$-polarized states, which would require terms such as $R^{ijji}$ . In other words, any superposition of $x$-and $y$-polarized states will not contribute to ${\mathcal R}_{MA}$. Hence, an MA-TA experiment will not infer the dephasing rate $\Gamma_{\rm deph}$. Less obviously the MA-TA signal is similarly insensitive to the population relaxation rate $\Gamma_{pop}$. This is true for all contributions to the ${\mathcal R}_{MA}$ signal
${\mathcal R}_{MA}={\rm GSB}+{\rm SE}+2*{\rm ESA(2B_{2g})}+{\rm ESA (2A_{1g})}+ {\rm ESA(2B_{1g})}$, which are
 \begin{equation} \label{eq:GSB_TAMA}
 {\rm GSB}: \frac{16}{3} \mu^4 I_{GSB}(\omega_1,\omega_3) 
 \end{equation}
 \begin{equation}
 {\rm SE}:\frac{8}{3} \mu^4 I_{SE}(\omega_1,\omega_3) 
 \end{equation}
 \begin{equation} \label{eq:ESA_2B2g_TAMA}
 {\rm ESA (2B_{2g})}: - \frac{4}{3}\mu^4 I_{ESA}(\omega_1,\omega_3) 
 \end{equation}
 \begin{equation}
 {\rm ESA (2A_{1g})}: - \frac{4}{3}\mu^4 I_{ESA}(\omega_1,\omega_3)
 \end{equation}
 \begin{equation} \label{eq:ESA_2B1G_TAMA}
 {\rm ESA (2B_{1g})}: - \frac{4}{3}\mu^4 I_{ESA}(\omega_1,\omega_3) 
 \end{equation}

Here, $\mu$ represents the transition dipole moment. $I(\omega_1,\omega_3)$ stands for the lineshape function with $\omega_1$ ($\omega_3$) as the central frequency of the pump (probe) pulse.
Microscopic theory of lineshape function is given elsewhere,\cite{Perlik2017} in the present work we select special detection frequencies where the lineshape functions are simple. 
All the TA-signal contributions in eqs. \ref{eq:GSB_TAMA}-\ref{eq:ESA_2B1G_TAMA} are delay-time independent and keep the ratio ${\rm GSB}: {\rm SE}: {\rm ESA (2B_{2g})}: {\rm ESA (2A_{1g})}: {\rm ESA (2B_{1g})} = 4:2:-1:-1:-1$, corresponding to the “weights” of diagrams used by Farrow et al.\cite{Farrow2008}. 
The independence of the MA-TA signal on population relaxation and pump-probe delay in general results from the molecule's symmetry, dictating that the transition strength to (GSB) and from (SE, ESA) the first excited state manifold is the same for the two single excited states. It is thus irrelevant for the TA signal if relaxation within the first excited state manifold has occurred. Accordingly, we observe a static MA-TA signal (within the experimentally probed delay range) in Figure \ref{fig:par_ort}. 

We now turn to the discussion of the detection wavelength dependence of $r(t)$. In the present experiment we employ broadband pump pulses and frequency-dispersed detection. In the $\alpha$-region of Figure \ref{fig:decay_mult}(a), centered around $\omega_3=2\pi c/790\,nm$ detection frequency, all possible excitation pathways, \textit{i.e.} $GSB, SE$ and $ESA$ contribute. 
Following the expression for $r(t)$ given in \ref{eq:anisotropy}, this lead to the following detection wavelength-dependent expression for the time-dependent anisotropy signal $r(t)$:
\begin{equation} \label{eq:r_790}
\\r_{790 nm}(t)=\frac{6G(t)-2}{10}=(1/10)+(3/10)e^{- \Gamma_{pop} t }
\end{equation}

Remarkably, all decoherence terms $\mathcal {I}(t_2)$ cancel, meaning that the anisotropy decay in square symmetric molecules only reports on population relaxation within the first excitonic manifold. According to eq. \ref{eq:r_790}, we expect an anisotropy value of 0.4 at $t_2=0$, and a value of $r(t)=0.1$ for times after population relaxation. These expectations agree with the experimental findings depicted in Figure \ref{fig:decay_mult}(b), red curve.
In terms of a molecular picture, the anisotropy value of 0.4 shows that the pumped and probed transition dipole moments are parallel, while the r=1/10 value signifies that the initial excitation becomes isotropic in a plane at the time of probing.

Around 890 nm detection wavelength, marked as $\beta$ in Figure \ref{fig:decay_mult}(a), we set $ I_{GSB}=0$ and $I_{SE}=I_{ESA}=1$. For $\omega_3=2\pi c / 975\,nm$, the contributions to \ref{eq:anisotropy} add up to:
\begin{equation} \label{eq:r_890}
\\r_{890 nm}(t))= \frac{4-6G(t)}{10}=(1/10)-(3/10)e^{-\Gamma_{pop} t}
\end{equation}
The resulting anisotropy decay at 890 nm shows the same time-constant as at 790 nm but approaches the common long-time $r$-value of 0.1 from a $r=-0.2$ starting value. The data in Figure \ref{fig:decay_mult}(b), blue curve, cannot be analyzed for early delay values. In the time window between zero and 100 fs, the coherent artifact due to pulse overlap effects contributes more strongly at 890 nm than at 790 nm due to the lower signal level for the former detection wavelength, which disturbs the anisotropy measurements. \cite{Dobryakov2005} 

An interesting signal is obtained for the combination of all the ESA signals, $I_{ESA}=1$, excluding all other pathways, $I_{GSB}=I_{SE}=0$. Then the anisotropy decay reads as
\begin{equation}
r_{ESA}(t)=\frac{2+3{\mathcal I}(t) }{20}= (1/10)+(3/20)e^{-\Gamma_{\rm dech} t}
\label{eq:r1000}
\end{equation}

and decay exclusively with $\Gamma_{{\rm dech}}$. This means that $r_{ESA}(t)$ lets us determine the decoherence time. In combination with $r_{790 nm}(t)$ and $r_{890 nm}(t)$, both decaying with $\Gamma_{{\rm pop}}$, we can discuss the proposal of $\Gamma_{{\rm dech}}=\Gamma_{pop}$, made by Raghavan et al. \cite{Raghavan2000}.

Therefore, we now turn to the discussion of signal in the $\gamma$-region of Figure \ref{fig:decay_mult}(c), where we expect only ESA contributions $r_{980 nm}= r_{ESA}$. 
Unfortunately, the signal at this edge is very weak and dominated by the coherent artifact \cite{Dobryakov2005}. The TA signal for orthogonal excitation pulses, $\mathcal{R}_{\perp}$, still carries the required information, as can be seen by rearranging equation \ref{eq:anisotropy}: 
\[
{\mathcal R}_{\perp}={\mathcal R}_{MA,ESA} \left(1-r(t)\right)
\]
Here the decay still represents the coherence dynamics ${\mathcal I}(t)$:
\[
{\mathcal R}_{\perp}=\frac{-16\mu^4}{3} \left(\frac{6-{\mathcal I}(t)}{20}\right)=\frac{-16\mu^4}{3} \left\{ \frac{3}{10}- \frac{1}{20} e^{-\Gamma_{\rm deph} t} \right\}
\]

As we measure only a single relaxation timescale in the $\gamma$ region of Figure \ref{fig:decay_mult}(a), we can attribute the 60 fs decay to the dephasing time. We find the same timescale for $r_{790 nm}(t)$ and $r_{840 nm}(t)$, both of which are defined by population relaxation. This supports the arguments given by Raghavan et al. \cite{Raghavan2000}, that square symmetric systems show equality between population relaxation and decoherence rates: $\Gamma_{pop} = \Gamma_{\rm dech}$. In other words the relaxation is isotropic within the plane. 
Hence, no experimental way exists to discriminate between these two rates, except when the square symmetry is lifted. \cite{Galli1993}. 

An additional feature of interest in all spectral ranges is the signal oscillations due to vibrational wavepacket motion. We attribute them to a normal mode at 650 $cm^{-1}$, previously observed by Farrow et al. \cite{Farrow2008}

\section{Conclusion}

This paper presents an experiment for the broadband and shot-to-shot detection of TAA signals. To demonstrate the experimental capability to measure femtosecond anisotropy dynamics in their detection wavelength dependence, we chose 2,3-Naphthalocyanin as a well-studied test sample. The spectral dependence of the observed 60 fs decay dynamics following a sub 20 fs excitation agrees with expectations from theory. The broadband detection of our TAA setup allowed us to discuss regions of different signal contributions separately. From this discussion, we learn that the anisotropy decay dynamics is always mono-exponential. We relate this finding to the molecule's square symmetry, in which population and electronic coherences are expected to decay with the same time constant. \cite{Raghavan2000}

The main component is a Sagnac-Interferometer inserted into the pump beampath of a TA experiment. This straightforward upgrade paves the way from TA to TAA and, in the future, toward polarization-controlled, partially non-collinear 2D-ES. 

\begin{acknowledgments}
ACKNOWLEDGMENTS
M.B., L.M., E.T. and J.H. acknowledge
funding by the Deutsche Forschungsgemeinschaft (DFG,
German Research Foundation) TRR 325 (project B8) and 514636421. J.H. acknowledges funding by DFG
under Germany’s Excellence Strategy-EXC 2089/1-
390776260. 
M.B. acknowledges funding by the Evangelisches Studienwerk Villigst for a Ph.D. stipend. 
F.S. acknowledges support by the Czech
Science Foundation (GACR) through grant no. 22-26376S.
\end{acknowledgments}

\section*{Data Availability Statement}
The data that support the findings of
this study are openly available in
zenodo at
http://doi.org/10.5281/zenodo.14923316.

\bibliography{QuellenAnisotropy}

\end{document}